\documentclass[english,pra,a4paper,aps,
twocolumn,floats]{revtex4}
\usepackage[latin9]{inputenc}
\usepackage{amsmath}
\usepackage{amsthm}
\usepackage{amssymb}
\usepackage{graphicx}
\usepackage{xcolor}

\makeatletter
\theoremstyle{plain}
\newtheorem{thm}{\protect\theoremname}
 \theoremstyle{definition}
 \newtheorem{defn}[thm]{\protect\definitionname}
 \theoremstyle{plain}
 
 \theoremstyle{definition}

\usepackage{MnSymbol}
\usepackage{slashbox}
\usepackage{hyperref}

\makeatother

\usepackage{babel}
 \providecommand{\corollaryname}{Corollary}
 \providecommand{\definitionname}{Definition}
\providecommand{\theoremname}{Theorem}
  \providecommand{\examplename}{Example}

\begin{document}

\title{One-particle and two-particle visibilities in bipartite entangled Gaussian states}

\author{Danko Georgiev}

\affiliation{Institute for Advanced Study, 30 Vasilaki Papadopulu Str., Varna 9010, Bulgaria}

\email{danko.georgiev@mail.bg}

\author{Leon Bello}

\affiliation{Department of Physics and the Institute of Nanotechnology and Advanced Materials, Bar Ilan University,
Ramat Gan 5290002, Israel}

\email{lionbello@gmail.com}

\author{Avishy Carmi}

\affiliation{Center for Quantum Information Science and Technology \& Faculty of Engineering Sciences, Ben-Gurion
University of the Negev, Beersheba 8410501, Israel}

\email{avcarmi@bgu.ac.il}

\author{Eliahu Cohen}

\affiliation{Faculty of Engineering and the Institute of Nanotechnology and Advanced Materials, Bar Ilan University,
Ramat Gan 5290002, Israel}

\email{eliahu.cohen@biu.ac.il}

%\pacs{03.65.Ta, 03.65.Ca, 03.65.Ud}

\date{June 2, 2021}

\begin{abstract}
Complementarity between one-particle visibility and two-particle visibility in discrete systems can be extended to bipartite quantum-entangled Gaussian states implemented with continuous-variable quantum optics. The meaning of the two-particle visibility originally defined by Jaeger, Horne, Shimony, and Vaidman with the use of an indirect method that first corrects the two-particle probability distribution by adding and subtracting other distributions with varying degree of entanglement, however, deserves further analysis. Furthermore, the origin of complementarity between one-particle visibility and two-particle visibility is somewhat elusive and it is not entirely clear what is the best way to associate particular two-particle quantum observables with the two-particle visibility. Here, we develop a direct method for quantifying the two-particle visibility based on measurement of a pair of two-particle observables that are compatible with the measured pair of single-particle observables. For each of the two-particle observables from the pair the corresponding visibility is computed, after which the absolute difference of the latter pair of visibilities is considered as a redefinition of the two-particle visibility. Our approach reveals an underlying mathematical symmetry as it treats the two pairs of one-particle or two-particle observables on equal footing by formally identifying all four observable distributions as rotated marginal distributions of the original two-particle probability distribution. The complementarity relation between one-particle visibility and two-particle visibility obtained with the direct method is exact in the limit of infinite Gaussian precision where the entangled Gaussian state approaches an ideal Einstein--Podolsky--Rosen state.
The presented results demonstrate the theoretical utility of rotated marginal distributions for elucidating the nature of two-particle visibility and provide tools for the development of quantum applications employing continuous variables.
\end{abstract}

\maketitle

\section{Introduction}

The particle nature of quantum theory is inbuilt in the tensor product composition of Hilbert spaces for composite physical systems \cite{vonNeumann1932,Dirac1967,Baggott2020}. The composite tensor product Hilbert space allows for realization of quantum-entangled states that are superpositions of tensor products of basis vectors for individual quantum systems such that the resulting composite quantum probability amplitudes are not separable \cite{Horodecki2009}.
For studying quantum entanglement in continuous-variable quantum systems, we have chosen to focus on entangled systems of superposed Gaussians as a minimal toy model due to relatively straightforward analytic integration of the resulting two-dimensional quantum probability distributions. Furthermore, entangled Gaussian states are practical for implementation in quantum technologies because such states can be readily produced \cite{Fang2010}, reliably controlled \cite{Laurat2005}, and efficiently measured \cite{Eisert2003,Braunstein2005,Rendell2005,Adesso2007,Serafini2017}.

The presence of quantum entanglement in bipartite systems could be manifested in the form of varying degrees of visibility of quantum interference patterns of single quantum observables or in the form of correlations of observable outcomes for pairs of compatible quantum observables \cite{Greenberger1993,Paul2018,Afrin2019,Kaur2020}.
Motivated by the pioneering work by Jaeger, Horne, Shimony, and Vaidman \cite{Jaeger1993,Jaeger1995} on quantum complementarity of one-particle and two-particle interference in four-beam interferometric setups, we have undertaken a detailed investigation aimed at finding the origin of this reported complementarity and elucidating the meaning of one-particle and two-particle visibilities in the case of continuous variables.
Within the context of bipartite entangled Gaussian states, we have addressed the following questions:

First, what is two-particle visibility? Also tightly related to this first question, what are the mathematical techniques and corresponding physical operations to determine interference visibilities from available multidimensional probability distributions? Suppose that we are granted the ability to determine the upper and lower envelopes of any interference pattern with a negligible experimental error. Even within such an idealized scenario, the original definition of visibility given as a ratio between the difference and the sum of upper and lower envelopes is well defined only for one-dimensional distributions. Apparently, this original definition can still be applied if the multidimensional distribution is mathematically preprocessed to reduce the overall number of dimensions to one. However, the dimensional reduction can be performed in at least two alternative ways. One procedure corresponding to the creation of a conditional distribution is slicing of the multidimensional distribution along an axis. The second procedure corresponding to the creation of an unconditional distribution is marginalization of the multidimensional distribution along an axis (the two procedures will be described by exact mathematical expressions within the following sections). Previous works on the problem \cite{Jaeger1993,Jaeger1995,Peled2020} were focused on the former approach, i.e., application of slicing through a given multidimensional distribution followed by fixing the ensuing unwanted consequences using the so-called ``corrections'' of the original multidimensional distribution. Here, we present the advantages of the latter approach, i.e., marginalization as a direct method of finding the visibilities without any correction of the original multidimensional distribution.

Second, how can the two-particle visibility be measured? Also tightly related to this second question, what are the complementary quantum observables corresponding to the one-particle and two-particle interference visibilities? The original definition of two-particle visibility by Jaeger, Horne, Shimony, and Vaidman \cite{Jaeger1993,Jaeger1995} was given in terms of slicing through a ``corrected'' two-dimensional distribution, which was constructed by addition and subtraction of other two-dimensional distributions. The first technical issue is that the slicing produces conditional distributions, which means that the two-particle visibility is expressed in some form of interdependence of a pair of observables where one of the two observables is postselected to a specific value. The second technical issue is that the correction of the two-dimensional quantum distribution may not guarantee the existence of a single quantum observable whose observable distribution is used to calculate the two-particle visibility for varying strength of entanglement; i.e., since the ``correction'' varies depending on the entanglement strength, it is not immediately clear why the sought-after single quantum observable cannot also vary as the entanglement varies. Here, we explicitly identify a pair of two-particle observables whose measurement is utilized to determine the two-particle visibility. This fact manifests a mathematical symmetry with the observation that single-particle visibilities are determined from a pair of single-particle observables.

Third, what is the origin and the physical mechanism that generates complementarity between the one-particle and two-particle interference visibilities? In single quantum systems, it is well known that quantum complementarity is due to the uncertainty relations between mutually unbiased observables acting on their Hilbert space \cite{Massar2008,Bagchi2016,Qureshi2013}. In bipartite quantum systems, however, the tensor product composition of Hilbert spaces allows for the existence of quantum-entangled states whose measurement allows for extraction of useful information about one of the systems by measuring the other system \cite{Einstein1935}.
Here, we show that in different bases the composite bipartite quantum state can always be decomposed in two complementary ways: either into a superposition of separable states or into a superposition of maximally entangled states. Noteworthy, these two complementary decompositions also display clearly as variables the sought-after one-particle and two-particle observables that are used for evaluating the one-particle and two-particle visibilities. The complementarity originates from an existing $\frac{\pi}{4}$ shift in the trigonometric functions appearing in the two decompositions.

The outline of the present work is as follows: In Section~\ref{sec:2}, we introduce the bipartite partially entangled Gaussian state, which is used for studying one-particle and two-particle visibility.
Furthermore, for different bases we present pairs of complementary decompositions, either in separable states or in maximally entangled states, which clearly pinpoint the origin of complementarity between one-particle and two-particle interference.
Next, in Section~\ref{sec:3}, we introduce the concepts of compatible (commuting) one-particle and two-particle observables, discuss their formal relation to
marginalization over a rotated axis with resulting rotated marginal distributions,
and derive a complementarity relation for symmetric setups.
Then, in Section~\ref{sec:4}, we generalize the complementarity relation between one-particle and two-particle observables for asymmetric setups.
In Section~\ref{sec:6}, we present another quantum complementarity relation involving incompatible (noncommuting) measurements for estimation of the one-particle visibility and the correlation between positions in the slits of the two entangled particles. Finally, we conclude with a discussion of the main findings and their significance. The meaning of essential technical jargon is clarified in the Appendixes.

\section{Partially entangled Gaussian state in different bases}
\label{sec:2}

\begin{figure*}[t]
\begin{centering}
\includegraphics[width=100mm]{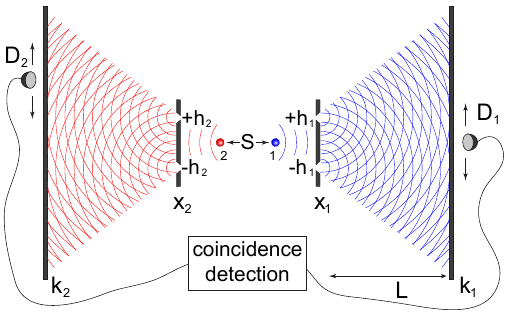}
\par\end{centering}
\caption{\label{fig:1}Paired double-slit experiment with entangled quanta. A source $S$ emits pairs of entangled quanta, each of which passes through a double slit. At the double slits, the pair of entangled quanta 1 and 2 are in the composite quantum state \eqref{eq:1} parametrized by the entanglement parameter $\xi$. Two particle detectors $D_1$ and $D_2$ feed forward their inputs to a coincidence detector for the recording of joint probability distributions. If both $D_1$ and $D_2$ are operated far away from the slits, they record Fraunhofer diffraction patterns which reveal the joint particle wavenumber distribution $P(k_1,k_2)=\left|\psi (k_{1},k_{2})\right|^{2}$. Alternatively, if either one of the two detectors or both of them are operated at the plane of the slits, they are able to record different, mutually incompatible joint probability distributions such as $P(x_1,k_2)=\left|\psi (x_{1},k_{2})\right|^{2}$, $P(k_1,x_2)=\left|\psi (k_{1},x_{2})\right|^{2}$, or $P(x_1,x_2)=\left|\psi (x_{1},x_{2})\right|^{2}$.
The coincidence detector ensures that the analyzed quanta are of the form \eqref{eq:1}; i.e., those single quanta that pass through the slits but whose entangled partner hits the opposite slit walls are excluded from the analysis. The $k$ distributions of quanta with wavelength $\lambda$ are assumed to be extracted, e.g., by scaled position measurements $2 \pi x / \lambda L$ \cite[\S~11.3.3]{Hecht2002} that are at distance $L$ sufficiently far away from the double slit so that the Fraunhofer diffraction pattern is obtained. An alternative practical way to extract the $k$ distributions is to record the position distribution from the focal plane of a lens that is focused onto the double slits \cite{Neves2007,Taguchi2008}. For our present purposes, we take for granted that the experimental realization of the Fourier transform of the position wave function is exact and the $k$ distributions can be measured with negligible experimental errors. The main research question that we address concerns what we do next after we have recorded $P(k_1,k_2)$.
}
\end{figure*}

Throughout this work, we will study the geometric structure of a partially entangled Gaussian state $\psi$ that can be utilized for the creation of one-particle and two-particle interference patterns.
One possible physical realization of such a state is through entangled photons in a paired double-slit setup \cite{Greenberger1993,Neves2007,Taguchi2008,Paul2018,Afrin2019,Kaur2020} (Fig.~\ref{fig:1}).
In the position basis, the partially entangled Gaussian state can be written as a superposition of maximally correlated and anticorrelated terms \cite{Peled2020}
\begin{widetext}
\begin{align}
\psi (x_{1},x_{2}) & =\sqrt{\frac{a}{2\pi}}B\Bigg[\left(e^{-a\left(x_{1}-h_{1}\right)^{2}}e^{-a\left(x_{2}-h_{2}\right)^{2}}+e^{-a\left(x_{1}+h_{1}\right)^{2}}e^{-a\left(x_{2}+h_{2}\right)^{2}}\right)\cos\left(\frac{\pi}{4}-\xi\right)
\nonumber \\
& \qquad\qquad
+\left(e^{-a\left(x_{1}-h_{1}\right)^{2}}e^{-a\left(x_{2}+h_{2}\right)^{2}}+e^{-a\left(x_{1}+h_{1}\right)^{2}}e^{-a\left(x_{2}-h_{2}\right)^{2}}\right)\sin\left(\frac{\pi}{4}-\xi\right)\Bigg]  \label{eq:1}\\
& =\sqrt{\frac{2a}{\pi}}Be^{-a(h_{1}^{2}+h_{2}^{2}+x_{1}^{2}+x_{2}^{2})}\left[\cosh(2ah_{1}x_{1}+2ah_{2}x_{2})\cos\left(\frac{\pi}{4}-\xi\right)+\cosh(2ah_{1}x_{1}-2ah_{2}x_{2})\sin\left(\frac{\pi}{4}-\xi\right)\right], \nonumber
\end{align}
\end{widetext}
where $a=\frac{1}{4\sigma^{2}}$ is a parameter that controls the precision of an individual Gaussian state (in statistics, the precision~$\frac{1}{\sigma^2}$ is the reciprocal of the variance $\sigma^2$),
$\pm h_{1},\pm h_{2}$ are the centers of the individual Gaussians,
\begin{equation}
B^{2}=\frac{e^{a\left(h_{1}^{2}+h_{2}^{2}\right)}} {\cosh\left[a\left(h_{1}^{2}+h_{2}^{2}\right)\right]+\cosh\left[a\left(h_{1}^{2}-h_{2}^{2}\right)\right]\cos\left(2\xi\right)},
\end{equation}
and $\xi$ is a parameter that controls the entanglement such that for $\xi=  0 + n \frac{\pi}{2}$, $n \in\mathbb{Z}$,
the state is separable, for $\xi= \frac{\pi}{4} + n \pi $ the state
is maximally correlated, and for $\xi= \frac{3\pi}{4} + n \pi $ the state
is maximally anticorrelated. Note that if the state at the second double slit has
a different Gaussian precision parameter $b$, we can always define new variables
$x_{2}=\bar{x}_{2}\sqrt{\frac{a}{b}}$ and $h_{2}=\bar{h}_{2}\sqrt{\frac{a}{b}}$
, which transform the state into the form \eqref{eq:1}. In other words, increasing the individual Gaussian precision of the wave function or rescaling the slits has the same effect.

To gain an alternative geometric insight into the structure of \eqref{eq:1}, we can use trigonometric angle addition identities for $\frac{\pi}{4}-\xi$
to rewrite the state as a superposition of two separable terms, one with four Gaussian peaks that have the same sign and one with four Gaussian peaks that have an opposite sign across the diagonal:
\begin{widetext}
\begin{equation}
\psi(x_{1},x_{2})=2\sqrt{\frac{a}{\pi}}Be^{-a(h_{1}^{2}+h_{2}^{2}+x_{1}^{2}+x_{2}^{2})}\left[\cosh(2ah_{1}x_{1})\cosh(2ah_{2}x_{2})\cos\xi+\sinh(2ah_{1}x_{1})\sinh(2ah_{2}x_{2})\sin\xi\right] . \label{eq:3}
\end{equation}
\end{widetext}
Eq.~\eqref{eq:3} is not a redundant decomposition of \eqref{eq:1}, but a complementary one. Even though the $x_1,x_2$ basis is used in both cases, \eqref{eq:1} is a decomposition into a superposition of maximally entangled states, whereas \eqref{eq:3} is a decomposition into a superposition of separable states. It will become clear in the subsequent mathematical derivations that the complementarity relation between one-particle and two-particle visibility originates exactly from the $\frac{\pi}{4}$ phase shift appearing in the separable versus the maximally entangled decomposition.

Fourier transform of \eqref{eq:1} gives the partially entangled wave function in wavenumber basis as a superposition of maximally correlated and anticorrelated terms:
\begin{widetext}
\begin{align}
\psi (k_{1},k_{2})& =\frac{1}{\sqrt{8a\pi}}Be^{-\frac{k_{1}^{2}+k_{2}^{2}}{4a}}\left[\left(e^{-\imath h_{1}k_{1}}e^{-\imath h_{2}k_{2}}+e^{\imath h_{1}k_{1}}e^{\imath h_{2}k_{2}}\right)\cos\left(\frac{\pi}{4}-\xi\right)
+\left(e^{-\imath h_{1}k_{1}}e^{\imath h_{2}k_{2}}+e^{\imath h_{1}k_{1}}e^{-\imath h_{2}k_{2}}\right)\sin\left(\frac{\pi}{4}-\xi\right)\right]\nonumber\\
&=\frac{1}{\sqrt{2a\pi}}Be^{-\frac{k_{1}^{2}+k_{2}^{2}}{4a}}\left[\cos\left(h_{1}k_{1}+h_{2}k_{2}\right)\cos\left(\frac{\pi}{4}-\xi\right)+\cos\left(h_{1}k_{1}-h_{2}k_{2}\right)\sin\left(\frac{\pi}{4}-\xi\right)\right] .
\label{eq:psi-k-theta}
\end{align}
\end{widetext}

The structure of \eqref{eq:psi-k-theta} could be further elucidated
by using trigonometric angle addition identities to rewrite
the state as a superposition of two separable states, one that is a product of fringes and one that is a product of antifringes:
\begin{widetext}
\begin{align}
\psi (k_{1},k_{2}) &=\frac{1}{4\sqrt{a\pi}}Be^{-\frac{k_{1}^{2}+k_{2}^{2}}{4a}}\left[\left(e^{-\imath h_{1}k_{1}}+e^{\imath h_{1}k_{1}}\right)\left(e^{-\imath h_{2}k_{2}}+e^{\imath h_{2}k_{2}}\right)\cos\xi
+\left(e^{-\imath h_{1}k_{1}}-e^{\imath h_{1}k_{1}}\right)\left(e^{-\imath h_{2}k_{2}}-e^{\imath h_{2}k_{2}}\right)\sin\xi\right]\nonumber\\
&=\frac{1}{\sqrt{a\pi}}B e^{-\frac{k_{1}^{2}+k_{2}^{2}}{4a}}\left[\cos\left(h_{1}k_{1}\right)\cos\left(h_{2}k_{2}\right)\cos\xi-\sin\left(h_{1}k_{1}\right)\sin\left(h_{2}k_{2}\right)\sin\xi\right] . \label{eq:ent-xi}
\end{align}
\end{widetext}
It can be seen that for $\xi= 0 + n \frac{\pi}{2}$ the state is separable,
whereas for $\xi = \frac{\pi}{4} + n \frac{\pi}{2} $
the state is maximally entangled. The cosine terms correspond to fringes, i.e., $\left(e^{-\imath h_{1}k_{1}}+e^{\imath h_{1}k_{1}}\right)=2\cos\left(h_{1}k_{1}\right)$
and $\left(e^{-\imath h_{2}k_{2}}+e^{\imath h_{2}k_{2}}\right)=2\cos\left(h_{2}k_{2}\right)$,
whereas the sine terms correspond to antifringes, i.e., $\left(e^{-\imath h_{1}k_{1}}-e^{\imath h_{1}k_{1}}\right)=-2\imath\sin\left(h_{1}k_{1}\right)$
and $\left(e^{-\imath h_{2}k_{2}}-e^{\imath h_{2}k_{2}}\right)=-2\imath\sin\left(h_{2}k_{2}\right)$.

A number of quantum complementarity relations constrain one-particle
visibility and two-particle visibility for discrete variables \cite{Jaeger1993,Jaeger1995,Hill1997,Wootters1998,Abouraddy2001}.
The previously used indirect method for assessment of two-particle visibility, however, is somewhat involved
because it requires a ``correction'' of $\left|\psi (k_{1},k_{2})\right|^{2}$ by addition and subtraction of two other terms \cite{Jaeger1993,Jaeger1995,Peled2020}
(for details on the original method proposed by Jaeger, Horne, Shimony, and Vaidman, see Appendix~\ref{sec:5}).
Here, our goal is to develop a direct method to quantify two-particle
visibility using only $\left|\psi (k_{1},k_{2})\right|^{2}$. We will also require that the complementarity is exact in the limit of infinite Gaussian precision $a\to\infty$ for every $\xi$ and all measured quantum observables (single-particle and two-particle observables) are treated on equal footing. In the exposition that follows, we will demonstrate that indeed such a direct method exists and it is based on
marginalization over rotated axes
of $\left|\psi (k_{1},k_{2})\right|^{2}$ [to be explicitly defined in Eq.~\eqref{eq:rmd} below and elaborated upon in Appendix~\ref{app}]. In outline,
two marginalizations
will give probability distributions for single-particle observables from which is determined the single-particle visibility, and
two other rotated marginalizations
will give probability distributions for two-particle observables from which is determined the two-particle visibility.
Importantly, all measured quantum observables are compatible, i.e., simultaneously measurable in the same experimental setting, as they commute with each other. This is noteworthy since quantum complementarity has been usually considered for incompatible observables, such as position and momentum of a single particle, which do not commute with each other and cannot be measured simultaneously in the same experimental setting.

\section{Special quantum complementarity relation for symmetric setups}
\label{sec:3}

For symmetric setups $h_{1}=h_{2}= h$, the joint probability distribution $P(k_1,k_2)=\left|\psi (k_{1},k_{2})\right|^{2}$ exhibits different geometric features for different values of the entanglement parameter $\xi$.
For $\xi=0$ the state of the two particles is separable into a product of fringes, whereas for $\xi=\frac{\pi}{2}$ the state is separable into a product of antifringes. The characteristic geometric feature of separable states is that they exhibit grooves and unit visibility in two perpendicular directions aligned with the $k_1$ and $k_2$ axes.
In contrast, the maximally entangled states exhibit grooves and unit visibility at only one of the two diagonal axes $k_{\pm}=\frac{1}{\sqrt{2}}\left(k_{1}\pm k_{2}\right)$.
For $\xi=\frac{\pi}{4}$, the maximally correlated state exhibits fringes only along the $k_{+}$~axis, whereas for $\xi=\frac{3\pi}{4}$ the maximally anticorrelated state exhibits fringes only along the $k_{-}$~axis.
Thus, the domain of the entanglement parameter $\xi$ extends in the interval $[0,\pi)$ before the period repeats.

Motivated by the characteristic geometry of maximally entangled states,
next we quantify the two-particle visibility using the marginal distributions
for $k_{+}$ and $k_{-}$. The marginal distributions for the standard
$k_{1}$,$k_{2}$ basis or the diagonal $k_{+}$,$k_{-}$ basis have
the physical meaning of performing measurements and extracting statistics
without accounting for the particular value obtained for the second
variable of the basis set, namely, $P(k_{1})=\int_{-\infty}^{\infty}\left|\psi (k_{1},k_{2})\right|^{2}dk_{2}$,
$P(k_{2})=\int_{-\infty}^{\infty}\left|\psi (k_{1},k_{2})\right|^{2}dk_{1}$,
$P(k_{+})=\int_{-\infty}^{\infty}\left|\psi (k_{+},k_{-})\right|^{2}dk_{-}$,
and $P(k_{-})=\int_{-\infty}^{\infty}\left|\psi (k_{+},k_{-})\right|^{2}dk_{+}$.
Formally, each rotated marginal distribution could be written as \cite{Temme1987,Deans1983}
\begin{widetext}
\begin{equation}
P(k_{s,\varphi})=\int_{-\infty}^{\infty}\int_{-\infty}^{\infty}\left|\psi (k_{1},k_{2})\right|^{2}\delta\left(k_{s,\varphi}-k_{1}\cos\varphi-k_{2}\sin\varphi\right)dk_{1}dk_{2}
\label{eq:rmd}
\end{equation}
\end{widetext}
as follows: $P(k_{1})\equiv P(k_{s,\varphi=0})$, $P(k_{2})\equiv P(k_{s,\varphi=\frac{\pi}{2}})$,
$P(k_{+})\equiv P(k_{s,\varphi=\frac{\pi}{4}})$ and $P(k_{-})\equiv P(k_{s,\varphi=-\frac{\pi}{4}})$.
It is worth emphasizing that we treat $\varphi$ as being fixed to a specific value thereby having only a single remaining free variable. For example, the $k_s$ axis rotated at $\varphi=\frac{\pi}{4}$ inside $k_1,k_2$ space coincides with the $k_+$ axis, hence we write $k_+\equiv k_{s,\varphi=\frac{\pi}{4}}$. In other words, the subscript ${\varphi=\frac{\pi}{4}}$ is intended as a reminder of the geometric interpretation of the $k_+$ axis as the particular axis that is rotated at this specified angle. This apparently cumbersome notation will prove to be useful in Section~\ref{sec:4} where we generalize the complementarity relation for asymmetric setups with two-particle observables that differ from $k_\pm$.

The visibility of an interference pattern in one-dimensional probability distribution $P(k)$ is usually defined as the ratio of the difference and sum of two smooth nonoscillatory functions $\textrm{env}^{+}(k)$ and  $\textrm{env}^{-}(k)$ referred to as upper and lower envelopes, respectively, which enclose tightly the oscillations of $P(k)$ from top and bottom:
\begin{equation}
\mathcal{V}(k)= \frac{\textrm{env}^{+}(k)-\textrm{env}^{-}(k)}{\textrm{env}^{+}(k)+\textrm{env}^{-}(k)} .
\end{equation}
Since probabilities are non-negative, both envelopes are also non-negative and the visibility $\mathcal{V}$ is bounded in the closed interval [0,1]. Typically, the visibility $\mathcal{V}(k)$ computed from $P(k)$ is not an explicit function of $k$ due to cancellation of the functional dependence on $k$ in the numerator and denominator of the fraction (Appendix~\ref{app-2}). Computing the visibility of interference patterns in multidimensional probability distributions, however, is not straightforward because slicing or marginalization along different rotated axes return, in general, different values of $\mathcal{V}$ as we see next.

After explicit integration of \eqref{eq:rmd} for different values of $\varphi$,
we obtain the following probability distributions:
\begin{widetext}
\begin{equation}
P(k_{1})  =\frac{B^{2}e^{-\frac{k_{1}^{2}}{2a}}}{2\sqrt{2a\pi}}\left\{ e^{-2ah_{2}^{2}}\left[\cos\left(2h_{1}k_{1}\right)+\cos\left(2\xi\right)\right]+1+\cos\left(2h_{1}k_{1}\right)\cos\left(2\xi\right)\right\} \label{eq:Pk1}
\end{equation}
with envelopes $\textrm{env}{}^{-}(k_1)$ obtained by setting $h_1 k_1\to\frac{\pi}{2}$, and
$\textrm{env}{}^{+}(k_1)$ obtained by setting $h_1 k_1\to\pi$.
Noteworthy, to obtain the correct envelopes all indicated substitutions should be performed only within the trigonometric functions leaving the leading amplitude intact. For details on envelope fitting based on some possible empirical data, see Appendix~\ref{app-2}.
\begin{equation}
P(k_{2}) =\frac{B^{2}e^{-\frac{k_{2}^{2}}{2a}}}{2\sqrt{2a\pi}}\left\{ e^{-2ah_{1}^{2}}\left[\cos\left(2h_{2}k_{2}\right)+\cos\left(2\xi\right)\right]+1+\cos\left(2h_{2}k_{2}\right)\cos\left(2\xi\right)\right\} \label{eq:Pk2}
\end{equation}
with envelopes $\textrm{env}{}^{-}(k_2)$ obtained by setting $h_2 k_2\to\frac{\pi}{2}$, and
$\textrm{env}{}^{+}(k_2)$ obtained by setting $h_2 k_2\to\pi$;
\begin{align}P(k_{\pm}) & =\frac{B^{2}e^{-\frac{k_{\pm}^{2}}{2a}}}{4\sqrt{2a\pi}}\Bigg\{2+2\left[e^{-ah_{1}^{2}}\cos\left(\sqrt{2}h_{1}k_{\pm}\right)+e^{-ah_{2}^{2}}\cos\left(\sqrt{2}h_{2}k_{\pm}\right)\right]\cos\left(2\xi\right)\nonumber\\
 & \qquad\qquad+e^{-a(h_{1}+h_{2})^{2}}\cos\left[\sqrt{2}\left(h_{1}-h_{2}\right)k_{\pm}\right]\left[1\mp\sin\left(2\xi\right)\right]+e^{-a(h_{1}-h_{2})^{2}}\cos\left[\sqrt{2}\left(h_{1}+h_{2}\right)k_{\pm}\right]\left[1\pm\sin\left(2\xi\right)\right]\Bigg\}
\end{align}
\end{widetext}
with envelopes $\textrm{env}{}^{-}(k_{\pm})$  obtained by setting $\sqrt{2}h_{1}k_{\pm}\to\frac{\pi}{2}$ and $\sqrt{2}h_{2}k_{\pm}\to\frac{\pi}{2}$, and $\textrm{env}{}^{+}(k_{\pm})$ obtained by setting $\sqrt{2}h_{1}k_{\pm}\to2\pi$ and $\sqrt{2}h_{2}k_{\pm}\to2\pi$.

It is worth pointing out that for $h_1\geq 1$ and $h_2\geq 1$, the envelopes are poor approximations as $a\to 1$; however, they are excellent approximations in the regime $a\gg 1$, and become perfect in the limit $a\to \infty$.

After introduction of the absolute value, because the upper and lower envelopes may switch their roles for different values of $\xi$, we compute the four unconditional visibilities
\begin{equation}
\mathcal{V}\left(k_{1}\right)  = \left|\frac{e^{-2ah_{2}^{2}}+\cos\left(2\xi\right)}{1+e^{-2ah_{2}^{2}}\cos\left(2\xi\right)}\right|, \label{eq:Vk1}
\end{equation}
\begin{equation}
\mathcal{V}\left(k_{2}\right)  = \left|\frac{e^{-2ah_{1}^{2}}+\cos\left(2\xi\right)}{1+e^{-2ah_{1}^{2}}\cos\left(2\xi\right)}\right|,
\end{equation}
\begin{widetext}
\begin{equation}
\mathcal{V}\left(k_{\pm}\right)  =\left|\frac{\left(e^{-ah_{1}^{2}}+e^{-ah_{2}^{2}}\right)\cos\left(2\xi\right)+e^{-a(h_{1}-h_{2})^{2}}\left[1\pm\sin\left(2\xi\right)\right]}{2+\left(e^{-ah_{1}^{2}}+e^{-ah_{2}^{2}}\right)\cos\left(2\xi\right)+e^{-a(h_{1}+h_{2})^{2}}\left[1\mp\sin\left(2\xi\right)\right]}\right|.
\end{equation}
\end{widetext}

For the symmetric setup $h_{1}=h_{2}$, we can define the single-particle
visibility as
\begin{equation}
V=\max\left[\mathcal{V}\left(k_{1}\right),\mathcal{V}\left(k_{2}\right)\right]
\end{equation}
 and the two-particle visibility as
\begin{equation}
W=\left|\mathcal{V}\left(k_{+}\right)-\mathcal{V}\left(k_{-}\right)\right|.
\end{equation}
The apparently different definitions for single-particle and two-particle visibilities highlight the geometric origin of the two measures:
in the two-dimensional surface provided by $P(k_1,k_2)$, the separable states contain grooves in two perpendicular directions that cross each other, whereas maximally entangled states contain parallel grooves in only one direction.
Thus, the choice of rotated marginalizations to generate an algebraic expression for the observable geometric characteristics of maximally entangled states becomes intuitively understandable; namely, marginalization in the direction along the parallel grooves will produce a one-dimensional distribution with visible fringes, whereas marginalization along the direction perpendicular to the grooves will produce a one-dimensional distribution with no fringes.
The negative sign in the two-particle visibility also has a geometric origin, namely, depending on the nature of quantum interference the parallel grooves for maximally entangled states are exhibited in only one of two distinct directions, which alternate as $\xi$ changes in multiples of $\frac{\pi}{2}$. In contrast, the positive sign in the one-particle visibility indicates that the crossing grooves for separable states always occur in the same two directions given by the $k_1$ axis and $k_2$ axis.

For symmetric setups with highly entangled Gaussian states in the limit of infinite Gaussian precision
$a\to\infty$, we obtain the exact results $\lim_{a\to\infty}V(a)=\left|\cos\left(2\xi\right)\right|$
and $\lim_{a\to\infty}W(a)=\left|\sin\left(2\xi\right)\right|$.
Therefore, the single-particle visibility and the two-particle visibility obey the
complementarity relation
\begin{equation}
\lim_{a\to\infty}\left(V^{2}+W^{2}\right)=\cos^{2}\left(2\xi\right)+\sin^{2}\left(2\xi\right)=1 . \label{eq:21}
\end{equation}
A naive attempt to directly generalize \eqref{eq:21} to asymmetric setups immediately fails because $k_\pm$ are not the correct two-particle observables for extracting the two-particle visibility. We will address this problem next.

\begin{figure*}[t]
\begin{centering}
\includegraphics[width=140mm]{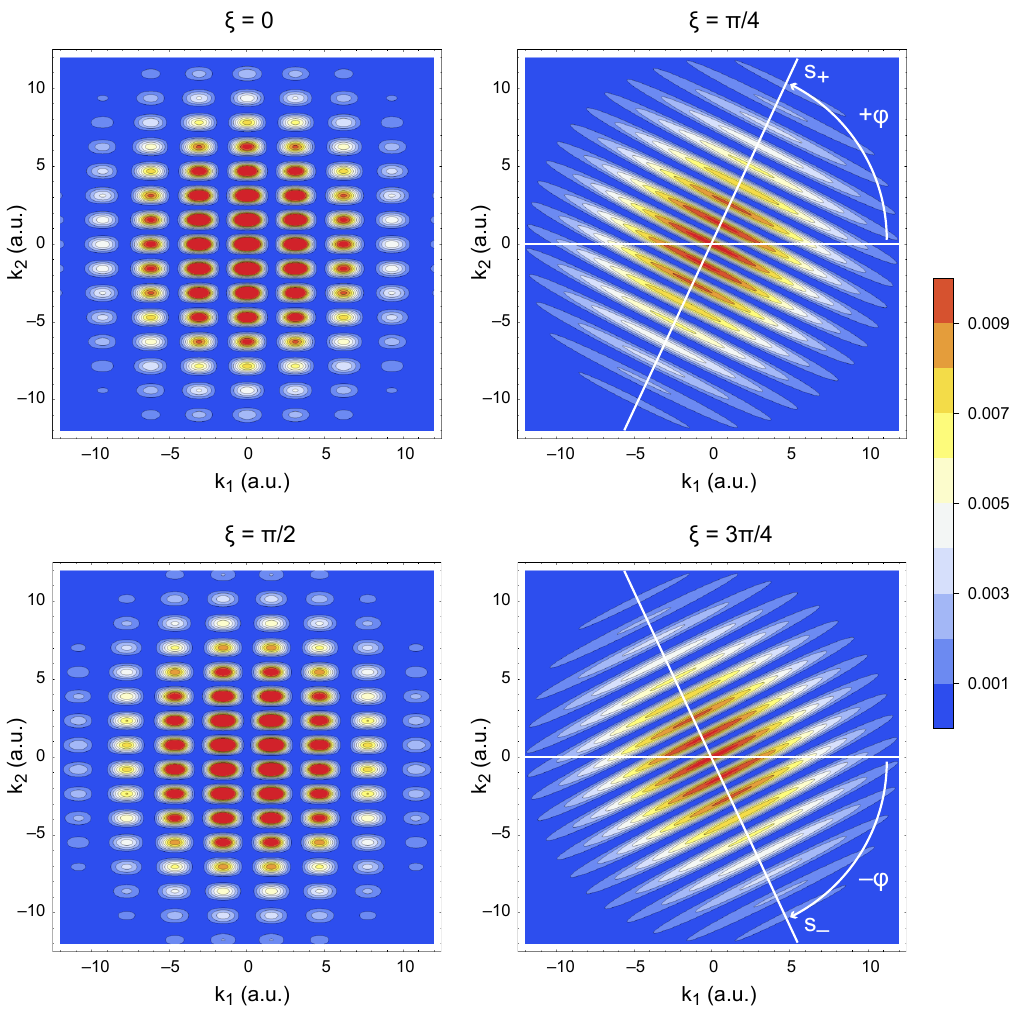}
\par\end{centering}
\caption{\label{fig:2}Joint probability distribution $P(k_1,k_2)=\left|\psi (k_{1},k_{2})\right|^{2}$
observed in Fraunhofer diffraction for asymmetric double slits $h_{1}=1,\,h_{2}=2$ with $a=30$ for different values of the entanglement parameter $\xi$. White lines indicate the generalized two-particle observables $s_\pm$ defined in \eqref{eq:s}. The wavenumbers are measured in arbitrary units (a.u.).}
\end{figure*}

\section{General quantum complementarity relation for asymmetric setups}
\label{sec:4}

For the asymmetric case $h_{1}\neq h_{2}$, the quantity $W$ no longer
provides a measure of two-particle visibility for two reasons: (1)
for the maximally entangled Gaussian states the
rotated marginal distributions
that exhibit perfect interference fringes are no longer located at an angle
of $\frac{\pi}{4}$ to one of the $k_{1},k_{2}$ axes, and (2) the
two relevant
rotated marginal distributions
are no longer perpendicular to each other (Fig.~\ref{fig:2}).
Taking into account the extra rotation introduced by
$h_{1}\neq h_{2}$, we can now consider two
rotated marginal distributions
at angles
\begin{equation}
\pm\varphi=\pm\arctan\left(\frac{h_{2}}{h_{1}}\right)
\end{equation}
with their associated visibilities $\mathcal{V}\left(s_{\pm}\right)\equiv\mathcal{V}\left(k_{s_{\pm},\pm\varphi}\right)$.
Thus, the general two-particle visibility is evaluated from the following generalized two-particle observables:
\begin{equation}
s_{\pm}	= \frac{h_{1}}{\sqrt{h_{1}^{2}+h_{2}^{2}}}k_{1} \pm \frac{h_{2}}{\sqrt{h_{1}^{2}+h_{2}^{2}}}k_{2} . \label{eq:s}
\end{equation}
The relevant two-particle observables are easy to guess from \eqref{eq:psi-k-theta}, where the geometric parameters of the paired double-slit setup are explicitly displayed. However, these two-particle observables could be determined from $P(k_1,k_2)$ alone with the use of multiple rotated marginal distributions and testing for which particular rotated axes $\pm\varphi$ the subsequent complementarity relations hold [see Eq.~\eqref{eq:VD} below].
After explicit integration of \eqref{eq:rmd} for the two axes located at $\pm\varphi$,
we obtain the following probability distributions:
\begin{widetext}
\begin{align}
P(s_{\pm}) & =\frac{B^{2}e^{-\frac{s_{\pm}^{2}}{2a}}}{4\sqrt{2a\pi}}\Bigg\{2+\cos\left(\frac{2s_{\pm}\left(h_{1}^{2}+h_{2}^{2}\right)}{\sqrt{h_{1}^{2}+h_{2}^{2}}}\right)
 \left[1 \pm \sin\left(2\xi\right)\right]+e^{-\frac{8ah_{1}^{2}h_{2}^{2}}{h_{1}^{2}+h_{2}^{2}}}\cos\left(\frac{2s_{\pm}\left(h_{1}^{2}-h_{2}^{2}\right)}{\sqrt{h_{1}^{2}+h_{2}^{2}}}\right)\left[1 \mp \sin\left(2\xi\right)\right]\nonumber \\
 & \qquad\qquad\quad+2e^{-\frac{2ah_{1}^{2}h_{2}^{2}}{h_{1}^{2}+h_{2}^{2}}}\left[\cos\left(\frac{2s_{\pm}h_{1}^{2}}{\sqrt{h_{1}^{2}+h_{2}^{2}}}\right)+\cos\left(\frac{2s_{\pm}h_{2}^{2}}{\sqrt{h_{1}^{2}+h_{2}^{2}}}\right)\right]\cos\left(2\xi\right)\Bigg\} .
\end{align}
\end{widetext}
The lower envelope $\textrm{env}{}^{-}(s_{\pm})$ is obtained by setting $\frac{s_{\pm}h_{1}^{2}}{\sqrt{h_{1}^{2}+h_{2}^{2}}}\to\frac{\pi}{4}$ and $\frac{s_{\pm}h_{2}^{2}}{\sqrt{h_{1}^{2}+h_{2}^{2}}}\to\frac{\pi}{4}$,
whereas the upper envelope $\textrm{env}{}^{+}(s_{\pm})$ is obtained by setting $\frac{s_{\pm}h_{1}^{2}}{\sqrt{h_{1}^{2}+h_{2}^{2}}}\to\pi$ and $\frac{s_{\pm}h_{2}^{2}}{\sqrt{h_{1}^{2}+h_{2}^{2}}}\to\pi$.
The corresponding visibilities are
\begin{equation}
\mathcal{V}\left(s_{\pm}\right)=\left|\frac{1+2e^{-\frac{2ah_{1}^{2}h_{2}^{2}}{h_{1}^{2}+h_{2}^{2}}}\cos\left(2\xi\right)\pm\sin\left(2\xi\right)}{2+2e^{-\frac{2ah_{1}^{2}h_{2}^{2}}{h_{1}^{2}+h_{2}^{2}}}\cos\left(2\xi\right)+e^{-\frac{8ah_{1}^{2}h_{2}^{2}}{h_{1}^{2}+h_{2}^{2}}}\left[1\mp\sin\left(2\xi\right)\right]}\right|.
\end{equation}
Thus, the two-particle visibility becomes
\begin{equation}
D=\left|\mathcal{V}\left(s_{+}\right)-\mathcal{V}\left(s_{-}\right)\right| . \label{eq:D}
\end{equation}
Consequently, for all bipartite double-slit setups (including asymmetric
ones) with highly entangled Gaussians in the limit $a\to\infty$,
we obtain the exact results $\lim_{a\to\infty}V(a)=\left|\cos\left(2\xi\right)\right|$
and $\lim_{a\to\infty}D(a)=\left|\sin\left(2\xi\right)\right|$. The
single-particle visibility and the two-particle visibility obey the
complementarity relation
\begin{equation}
\lim_{a\to\infty}\left(V^{2}+D^{2}\right)=\cos^{2}\left(2\xi\right)+\sin^{2}\left(2\xi\right)=1 .
\label{eq:VD}
\end{equation}
For symmetric setups $h_{1}=h_{2}$, we encounter the special case when $D=W$.

Because the quantum complementarity relation \eqref{eq:VD} is asymptotically
tight, for real-world quantum applications with finite $a$ it would
be helpful to have a measure for the deviation from unity,
\begin{equation}
\epsilon=1-V^{2}-D^{2} .
\end{equation}
With imposed conditions $h_{1}\geq1$, $h_{2}\geq1$ and $a\geq2$,
the deviation $\epsilon$ is bounded by
\begin{equation}
\left|\epsilon\right|<2e^{-\frac{2ah_{1}^{2}h_{2}^{2}}{h_{1}^{2}+h_{2}^{2}}} .\label{eq:bound}
\end{equation}
The observed oscillation around unity might be related to the aforementioned approximate nature of the computed envelopes, which become exact only in the limit of infinite Gaussian precision $a\to\infty$.

\section{Quantum complementarity for incompatible observables}
\label{sec:6}

\begin{figure*}[t]
\begin{centering}
\includegraphics[width=140mm]{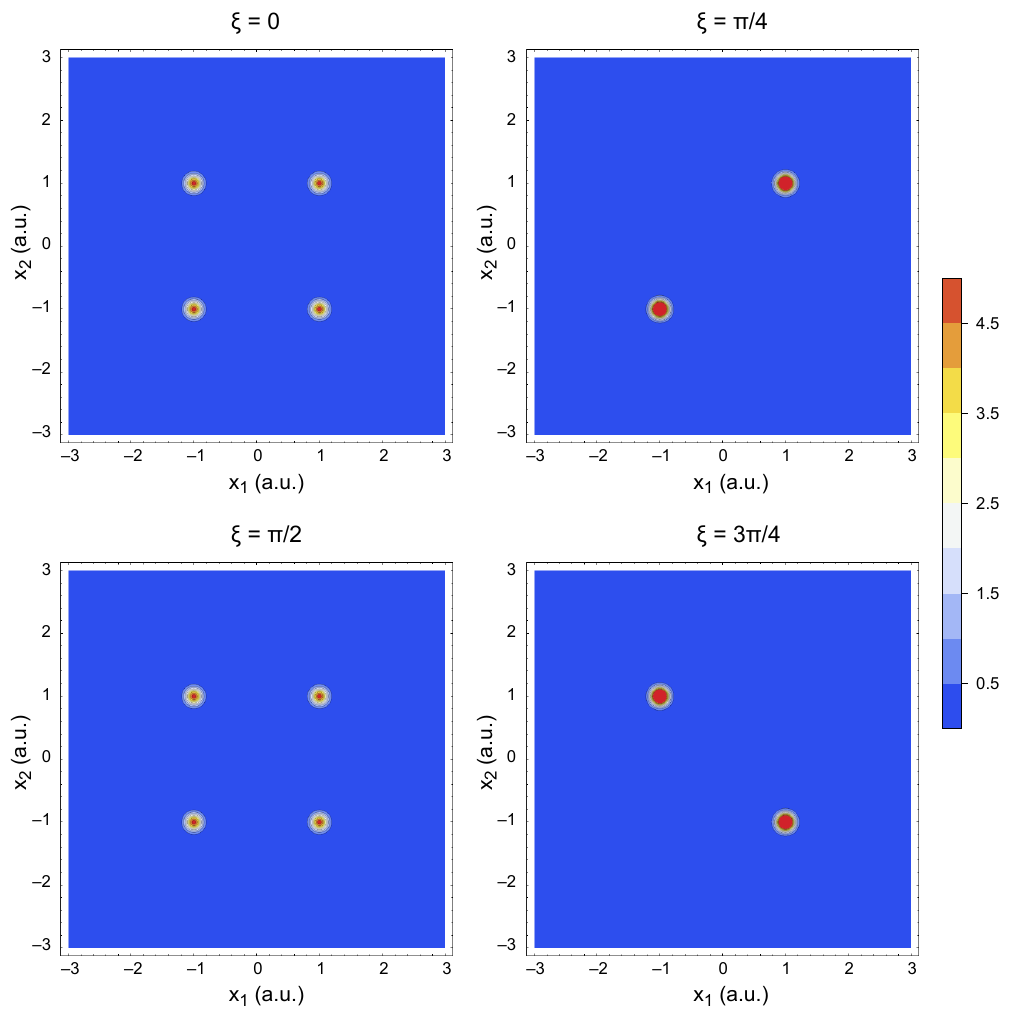}
\par\end{centering}
\caption{\label{fig:3}Joint probability distribution $P(x_1,x_2)=\left|\psi (x_{1},x_{2})\right|^{2}$ observed at the plane of the slits for symmetric double slits $h_{1}=h_{2}=1$ with $a=30$ for different values of the entanglement parameter $\xi$. The positions are measured in arbitrary units~(a.u.).}
\end{figure*}

\begin{figure*}[t]
\begin{centering}
\includegraphics[width=140mm]{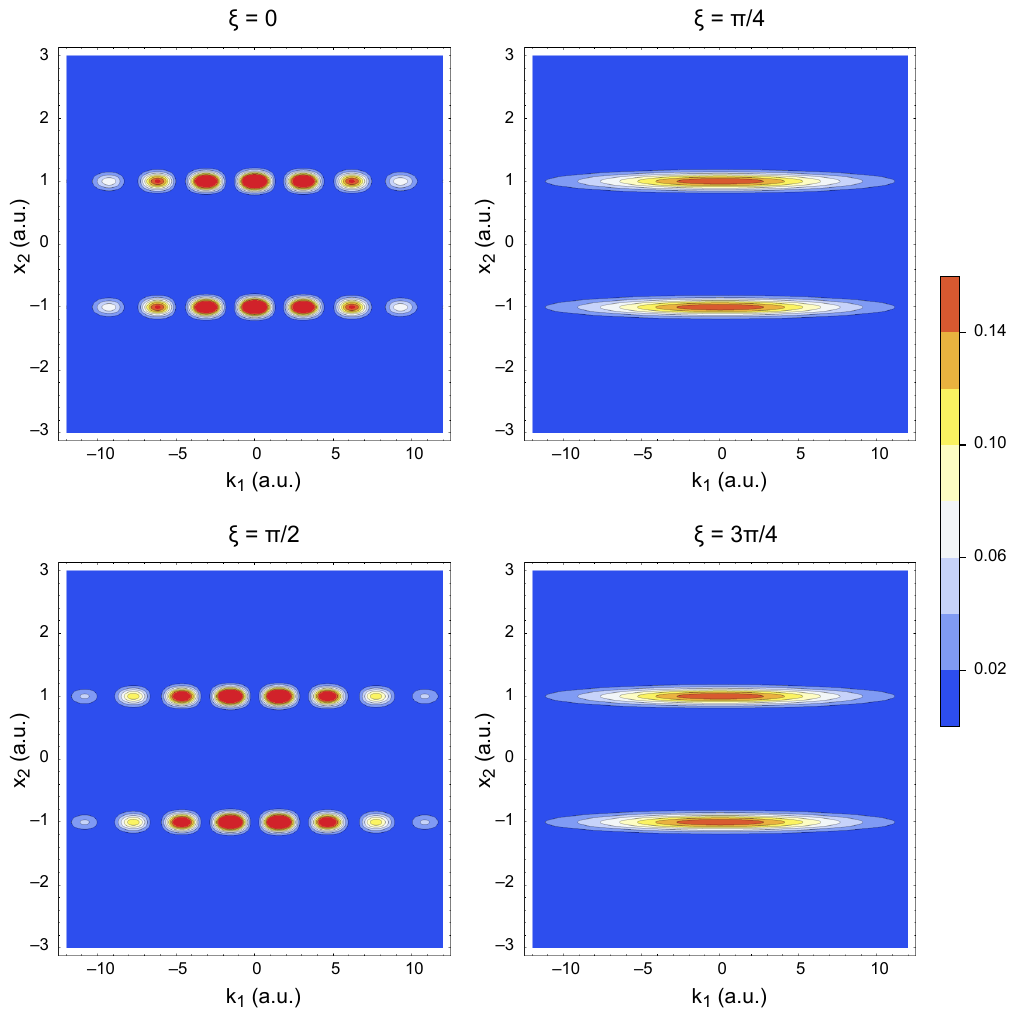}
\par\end{centering}
\caption{\label{fig:4}Joint probability distribution $P(k_1,x_2)=\left|\psi (k_{1},x_{2})\right|^{2}$ observed in Fraunhofer diffraction for the first slit and at the plane of the second slit for asymmetric double slits $h_{1}=h_{2}=1$ with $a=30$ for different values of the entanglement parameter~$\xi$.}
\end{figure*}

\begin{figure*}[t]
\begin{centering}
\includegraphics[width=160mm]{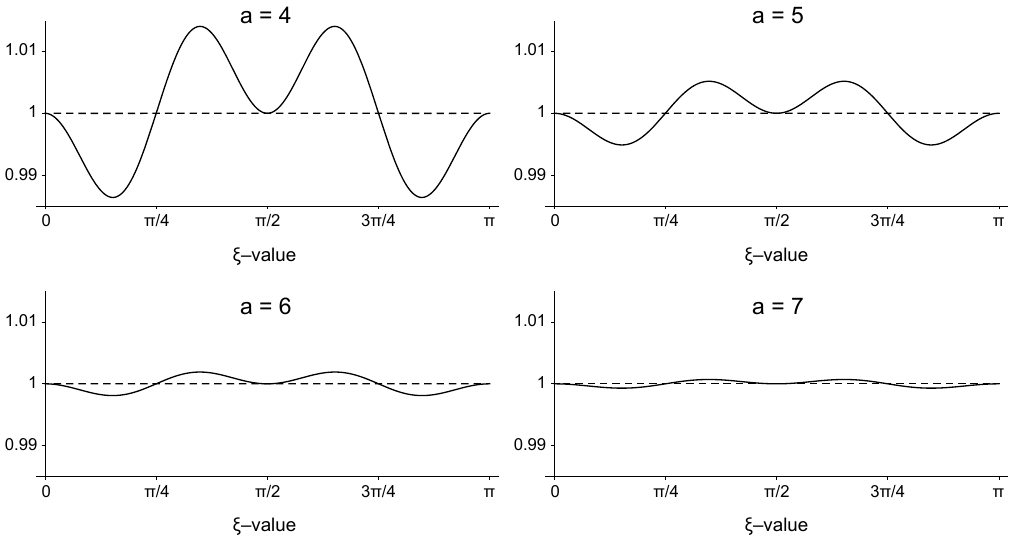}
\par\end{centering}
\caption{\label{fig:5}Comparison of $V^{2}+D^{2}$ (solid line) and $V^{2}+R^{2}$ (dashed line) for different values of the Gaussian precision parameter $a$ in symmetric paired double slit experiment with $h_{1}=h_{2}=1$.
The convergence to unity of the relation $V^{2}+R^{2}$ for incompatible (noncommuting) observables is faster compared with the relation $V^{2}+D^{2}$ for commuting observables.}
\end{figure*}

Previous research has demonstrated that probes located at the arms of a Mach--Zehnder interferometer are able to reduce the appearance of interference fringes at the interferometer exit depending on the ability of the probes to distinguish the two interferometer arms \cite{Massar2008,Qureshi2013,Zhang2015,Bagchi2016,Basso2020}. In the context of the partially entangled bipartite Gaussian state \eqref{eq:1}, the distinguishability could be computed from the Pearson correlation between the two position observables,
\begin{equation}
\varrho({x_{1},x_{2}})=\frac{\textrm{Cov}(x_{1},x_{2})}{\sqrt{\textrm{Var}(x_{1})\textrm{Var}(x_{2})}} , \label{eq:pearson-x}
\end{equation}
where $\textrm{Cov}(x_{1},x_{2})=\langle x_{1}x_{2}\rangle-\langle x_{1}\rangle\langle x_{2}\rangle$,
$\textrm{Var}(x_{1})=\textrm{Cov}(x_{1},x_{1})$,
$\textrm{Var}(x_{2})=\textrm{Cov}(x_{2},x_{2})$, and
\begin{align}
\langle x_{1}x_{2}\rangle	&=\int_{-\infty}^{+\infty}\int_{-\infty}^{+\infty}x_{1}x_{2}\left|\psi(x_{1},x_{2})\right|^{2}\,dx_{1}dx_{2} , \\
\langle x_{1}\rangle	&=\int_{-\infty}^{+\infty}\int_{-\infty}^{+\infty}x_{1}\left|\psi(x_{1},x_{2})\right|^{2}\,dx_{1}dx_{2} , \\
\langle x_{2}\rangle	&=\int_{-\infty}^{+\infty}\int_{-\infty}^{+\infty}x_{2}\left|\psi(x_{1},x_{2})\right|^{2}\,dx_{1}dx_{2}.
\end{align}
Explicit integration of $\left|\psi(x_{1},x_{2})\right|^{2}$ gives
\begin{equation}
\textrm{Cov}(x_{1},x_{2}) =\frac{B^{2}}{2}h_{1}h_{2}\sin(2\xi),
\end{equation}
\begin{widetext}
\begin{align}
\textrm{Var}(x_{1}) &=\frac{B^{2}}{8a}e^{-2a(h_{1}^{2}+h_{2}^{2})}\left\{ 1+e^{2ah_{2}^{2}}\cos(2\xi)+e^{2ah_{1}^{2}}\left[e^{2ah_{2}^{2}}+\cos(2\xi)\right]\left(1+4ah_{1}^{2}\right)\right\} , \\
\textrm{Var}(x_{2}) &=\frac{B^{2}}{8a}e^{-2a(h_{1}^{2}+h_{2}^{2})}\left\{ 1+e^{2ah_{1}^{2}}\cos(2\xi)+e^{2ah_{2}^{2}}\left[e^{2ah_{1}^{2}}+\cos(2\xi)\right]\left(1+4ah_{2}^{2}\right)\right\} .
\end{align}
\end{widetext}
For $\xi=\frac{\pi}{4}+n\frac{\pi}{2}$, the correlation or anticorrelation becomes perfect in the limit of infinite Gaussian precision $a\to\infty$, namely, $\lim_{a\to\infty} | \varrho(x_1,x_2)| = 1$. For any finite value of $a$, however, there will be a drop in the correlation due to the fact that the positions within the aperture of the slits are not correlated, i.e., that the individual Gaussian regions in Fig.~\ref{fig:3} have a nonzero extent. Because we are only interested in quantum interference across the two slits, but not in the quantum interference within each slit aperture, it is possible to normalize the correlation using the value for $\xi=\frac{\pi}{4}$ and use it as a measure of distinguishability of the two slits as follows:
\begin{equation}
R = \left|\frac{\varrho({x_{1},x_{2},\xi})}{\varrho({x_{1},x_{2},\xi=\frac{\pi}{4}})} \right| .
\end{equation}
Now, in order to see how the entanglement between the two systems affects the quantum interference of, say, the first system, we can measure the bipartite state in a mixed basis:
\begin{widetext}
\begin{equation}
\psi(k_{1},x_{2})=\sqrt{\frac{2}{\pi}}Be^{-\frac{k_{1}^{2}}{4a}}e^{-a(h_{2}^{2}+x_{2}^{2})}\left[\cos(h_{1}k_{1})\cosh(2ah_{2}x_{2})\cos\xi-\imath\sin(h_{1}k_{1})\sinh(2ah_{2}x_{2})\sin\xi\right] .
\end{equation}
The corresponding marginal distribution computed from $|\psi(k_{1},x_{2})|^2$ for the first system is
\begin{align}
P(k_{1}) =\int_{-\infty}^{\infty}\left|\psi(k_{1},x_{2})\right|^{2}dx_{2}
 =\frac{B^{2}e^{-\frac{k_{1}^{2}}{2a}}}{2\sqrt{2a\pi}} \left\{ e^{-2ah_{2}^{2}}\left[\cos(2h_{1}k_{1}) + \cos(2\xi)\right] + 1+\cos(2h_{1}k_{1})\cos(2\xi) \right\} . \label{eq:Pk1b}
\end{align}
\end{widetext}
Consistent with the no-communication theorem \cite{Eberhard1978,Eberhard1989,Ghirardi1980,Peres2004}, the latter distribution \eqref{eq:Pk1b} is equal to \eqref{eq:Pk1} obtained from marginalization of $|\psi(k_{1},k_{2})|^2$ and has the same visibility $V=\mathcal{V}(k_1)$ given by \eqref{eq:Vk1}.
Interference fringes in $|\psi(k_{1},x_{2})|^2$ are perfectly visible when the bipartite state is separable, $\xi=0+n\frac{\pi}{2}$, and are completely absent when the state is maximally entangled, $\xi=\frac{\pi}{4}+n\frac{\pi}{2}$ (Fig.~\ref{fig:4}). Combining $R^2$ and $V^2$ also gives a perfect quantum complementarity relation in the limit of infinite Gaussian precision,
\begin{equation}
\lim_{a\to\infty}\left(R^{2}+V^{2}\right)=\sin^{2}\left(2\xi\right)+\cos^{2}\left(2\xi\right)=1.
\label{eq:VR}
\end{equation}
The convergence to unity with respect to the Gaussian precision parameter $a$ of the relation $V^{2}+R^{2}$ for incompatible (noncommuting) observables is faster compared with the relation $V^{2}+D^{2}$ for commuting observables (Fig.~\ref{fig:5}). The drawback of the relation for incompatible observables is that $V$ and $R$ cannot be determined with a single setting of the measurement apparatus, but need two alternative settings for incompatible experimental measurements.

At this point, one might be interested in the possible use of correlation of outcomes in the wavenumber basis,
\begin{equation}
\varrho({k_{1},k_{2}})=\frac{\textrm{Cov}(k_{1},k_{2})}{\sqrt{\textrm{Var}(k_{1})\textrm{Var}(k_{2})}} ,
\end{equation}
for the construction of an alternative complementarity relation for commuting observables. Indeed from the covariance and individual variances
\begin{equation}
\textrm{Cov}(k_{1},k_{2}) =-2a^{2}B^{2}e^{-2a(h_{1}^{2}+h_{2}^{2})}h_{1}h_{2}\sin(2\xi) ,
\end{equation}
\begin{widetext}
\begin{align}
\textrm{Var}(k_{1}) &=\frac{1}{2}aB^{2}e^{-2ah_{2}^{2}}\left\{ e^{2ah_{2}^{2}}+\cos(2\xi)+e^{-2ah_{1}^{2}}\left[1+e^{2ah_{2}^{2}}\cos(2\xi)\right]\left(1-4ah_{1}^{2}\right)\right\} , \\
\textrm{Var}(k_{2}) &=\frac{1}{2}aB^{2}e^{-2ah_{1}^{2}}\left\{ e^{2ah_{1}^{2}}+\cos(2\xi)+e^{-2ah_{2}^{2}}\left[1+e^{2ah_{1}^{2}}\cos(2\xi)\right]\left(1-4ah_{2}^{2}\right)\right\} ,
\end{align}
\end{widetext}
one can create a normalized correlation measure
\begin{equation}
S = \left|\frac{\varrho({k_1,k_2,\xi})}{\varrho({k_{1},k_{2},\xi=\frac{\pi}{4}})} \right| \label{eq:53}
\end{equation}
for which
\begin{equation}
\lim_{a\to\infty}\left(S^{2}+V^{2}\right)=\sin^{2}\left(2\xi\right)+\cos^{2}\left(2\xi\right)=1 . \label{eq:54}
\end{equation}
Despite the superficial similarity with the other relations derived so far, there is a serious downside to formula \eqref{eq:54} which undermines its practical utility. Whereas the correlation in position  basis $|\varrho({x_1,x_2,\xi})|$ becomes unity in the limit of infinite Gaussian precision, $\lim_{a\to\infty}|\varrho({x_1,x_2,\xi})|=1$, the correlation in wavenumber basis $|\varrho({k_1,k_2,\xi})|$ vanishes in the limit of infinite Gaussian precision, $\lim_{a\to\infty}|\varrho({k_1,k_2,\xi})|=0$.
This means that if one replaces $R$ with $|\varrho({x_1,x_2,\xi})|$ in the complementarity relation, it will still converge to unity,
\begin{equation}
\lim_{a\to\infty}\left[\varrho^2({x_1,x_2,\xi})+V^{2}\right]=\sin^{2}\left(2\xi\right)+\cos^{2}\left(2\xi\right)=1,
\end{equation}
but if one replaces $S$ with $|\varrho({k_1,k_2,\xi})|$, the limit is changed:
\begin{equation}
\lim_{a\to\infty}\left[\varrho^2({k_1,k_2,\xi})+V^{2}\right]=\cos^{2}\left(2\xi\right) .
\end{equation}
In other words, any attempts to use \eqref{eq:54} will face the practical problem that sensitivity of measuring devices will be exceeded even for modest values of $a$. For example, in a symmetric setup with $h_1=h_2=1$ and $a=10$, the correlation is negligible $\varrho({k_{1},k_{2},\xi=\frac{\pi}{4}})\approx 3 \times 10^{-32}$. This tiny value needs to be measurable first before one is able to normalize the measured value according to \eqref{eq:53}. In essence, the complementarity relation \eqref{eq:54} is not practical from an experimental viewpoint.

\section{Discussion}

In this work, we have derived a complementarity relation \eqref{eq:VD} between one-particle visibility and two-particle visibility for bipartite partially entangled Gaussian states. This complementarity relation, obtained for continuous-variable systems, is reminiscent of a relation obtained for binary-outcome observables in interferometric setups \cite{Jaeger1993,Jaeger1995}.
There are several aspects, however, that differentiate our proposal \eqref{eq:VD} from earlier works \cite{Greenberger1988,Englert1992,Englert1996,Peled2020}.

First, we have brought to the forefront the fact that the complementarity relation between one-particle visibility and two-particle visibility is one involving only compatible (commuting) observables. This is particularly clear in our derivations because we work only with a single quantum probability distribution $P(k_1,k_2)$ without ``correcting it.''

Second, we have explicitly identified the pair of two-particle quantum observables whose visibilities are combined in order to produce the two-particle visibility \eqref{eq:D}. Previous research in two-particle visibility based on so-called corrected distribution \cite{Jaeger1993,Jaeger1995,Peled2020} did not treat the two-particle visibility with the same mathematical procedure as single-particle visibility, because the former was determined by conditional slicing through two-dimensional distribution, whereas the latter was determined from unconditional (marginal) one-dimensional distribution. Here, we employed only marginal distributions for both single-particle and two-particle observables, which restored the symmetry of the mathematical procedures and put the resulting visibilities on equal footing.

Third, we have shown that in the limit of infinite Gaussian precision, the bipartite quantum entanglement leads to manifested position correlation, $\lim_{a\to\infty}|\varrho(x_1,x_2)|=1$, but vanishing wavenumber correlation, $\lim_{a\to\infty}|\varrho(k_1,k_2)|=0$. From the former fact, one could easily construct noncommuting quantum complementarity relations for position and wavenumber of a single target particle. In particular, the stronger the position $x_1$ of the target particle is entangled with some observable (in this case the position $x_2$) of the second probe particle, the weaker the interference fringes visible in the wavenumber distribution $P(k_1)$ will be.
What is interesting, however, is that the strength of the quantum entanglement between the two particles can be extracted from the distribution $P(k_1,k_2)$ despite the fact that the correlation $|\varrho(k_1,k_2)|$ is vanishing. Our formula \eqref{eq:D} extracts the strength of quantum entanglement from the overall geometry of $P(k_1,k_2)$ through suitably chosen pairs of
rotated marginal distributions and computation of the resulting visibilities.

The presented results are limited to pure bipartite states. Consideration of mixed states is one possible way for generalizing the reported complementarity relation, which will be invariably converted into an equality. An alternative way is to consider purification of the mixed bipartite state using a third quantum system with appropriate dimensionality of the Hilbert space. In this latter approach, the exact complementarity relation to unity will be preserved; however, one will need to construct a generalized notion of $n$-particle visibility in which can be specified $n=3$. We leave such investigations for future work.

While the discussion in this work was presented in terms of a paired double-slit setup, it applies just as well to a continuous-variable description of quantum fields, and could be easily produced in a quantum optical setup. In that setup, the role of the particles' position and momentum can be assumed by the field quadrature amplitudes, and the partially entangled state may be implemented by a two-mode squeezed vacuum state. The pair of double slits is isomorphic to a pair of Mach--Zehnder interferometers (as in the famous Franson experiment \cite{Franson1989,Ou1990,Aerts1999}), where the distance between the slits is equivalent to the delay between the two-interferometer arms, and the interference pattern measured on the screen can be replaced by a homodyne measurement.

In summary, the presented results provide a geometric characterization of bipartite quantum entanglement using a basis in which the single-particle observables exhibit vanishing correlation. In such case, the information about the entanglement strength is stored in two-particle observables.
The existence of a complementarity relation in the wavenumber basis between one-particle observables and two-particle observables justifies their characterization as complementary observables even though they are compatible; i.e., $\hat{k}_1$ and $\hat{k}_2$ commute with each other and with any linear combination $a\hat{k}_1+b\hat{k}_2$, where $a,b\in\mathbb{R}$.
Direct comparison of relations \eqref{eq:VD} for compatible observables and \eqref{eq:VR} for incompatible observables shows that because the single-particle visibility $V$ is present in both of them, the two-particle visibility $D$ computed from the two-particle wavenumbers is able to provide indirect information about the position correlation $R$ of the two particles, and vice versa. In other words, measurement of $D$ reveals with certainty the value of $R$ (within a controllable error that vanishes in the limit of infinite Gaussian precision) that would have been obtained from measurement of the two-particle positions. Thus, the present operational approach towards extraction of two-particle visibility from appropriate two-particle observables may be also useful for the development of new protocols for quantum communication with continuous variables.

\section*{Acknowledgements}

We wish to thank two anonymous reviewers for very helpful comments.
This research was supported by Grant No.~FQXi-RFP-CPW-2006 from the Foundational Questions Institute and Fetzer Franklin Fund, a donor-advised fund of Silicon Valley Community Foundation, by the Israeli Innovation authority (Grants No.~70002 and No.~73795), by the Pazy foundation, and by the Quantum Science and Technology Program of the Israeli Council of Higher Education.

\appendix
%%%%%%%%%%%%%%%%%%%%%%%%%%%%%%%%%%%
%%%%%%%%%%%%%%%%%%%%%%%%%%%%%%%%%%%

\section{Indirect method based on ``corrected'' distribution}
\label{sec:5}

\begin{figure*}[t]
\begin{centering}
\includegraphics[width=140mm]{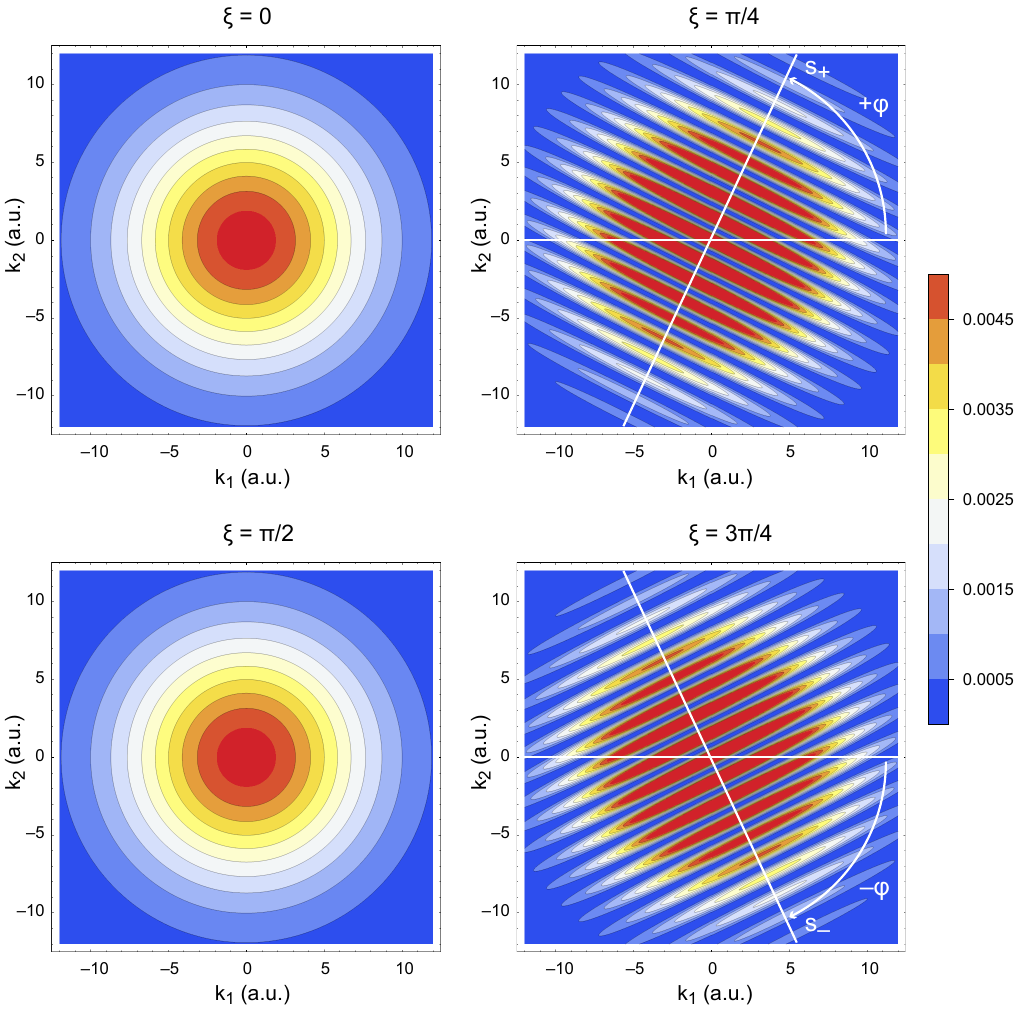}
\par\end{centering}
\caption{\label{fig:6}$\bar{P}(k_{1},k_{2})$ for asymmetric double slits $h_{1}=1,\,h_{2}=2$ with $a=30$ for different values of the entanglement parameter $\xi$.}
\end{figure*}

\begin{figure*}[t]
\begin{centering}
\includegraphics[width=160mm]{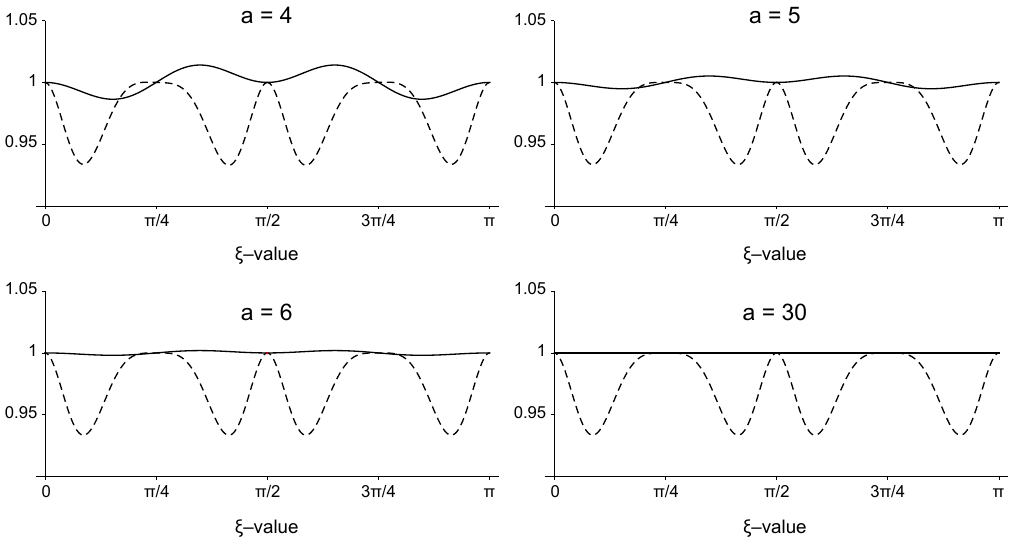}
\par\end{centering}
\caption{\label{fig:7}Comparison of $V^{2}+D^{2}$ (solid line) and $V^{2}+F^{2}$ (dashed line) for different values of the Gaussian precision parameter $a$ in the symmetric paired double-slit experiment with $h_{1}=h_{2}=1$.
The convergence to unity of $V^{2}+D^{2}$ is exponential with respect to the Gaussian precision parameter $a$ such
that the deviation $\left|\epsilon\right|$ is bounded by \eqref{eq:bound}. In contrast, $V^{2}+F^{2}$ does not converge to unity.}
\end{figure*}

The previously used indirect method for computation of the two-particle visibility relies on
somewhat involved addition and subtraction of distributions. To eliminate fringes in the $\xi=0$ case, the distribution
$P(k_{1},\xi)P(k_{2},\xi)$ is subtracted from $\left|\psi (k_{1},k_{2})\right|^{2}$.
To further correct occurrence of negative values, the distribution
$P(k_{1},\xi=\frac{\pi}{4})P(k_{2},\xi=\frac{\pi}{4})$ is added, resulting
in the following ``corrected'' distribution:
\begin{align}
\bar{P}(k_{1},k_{2}) &= \left|\psi (k_{1},k_{2})\right|^{2}-P(k_{1},\xi)P(k_{2},\xi) \nonumber\\
& \qquad +P(k_{1},\xi=\frac{\pi}{4})P(k_{2},\xi=\frac{\pi}{4}) . \label{eq:corrected}
\end{align}
For both symmetric or asymmetric setups, the separable cases $\xi=0 + n\frac{\pi}{2}$ contain no
interference fringes, whereas for the maximally entangled cases $\xi=\frac{\pi}{4} + n\frac{\pi}{2}$
perfect interference fringes are exhibited along one of the axes $s_{\pm}$
at angles $\pm\varphi=\pm\arctan\left(\frac{h_{2}}{h_{1}}\right)$ (Fig.~\ref{fig:6}).
The main motivation for introducing \eqref{eq:corrected} is that
the two-particle visibility could be computed using a slice of the
corrected distribution through the origin, i.e., by conditionally setting
the corresponding perpendicular variables to zero, $s_{+}^{\perp}=0$
or $s_{-}^{\perp}=0$. Since this method modifies the original distribution $\left|\psi (k_{1},k_{2})\right|^{2}$, it outputs results that differ from those obtained with the direct method based only on $\left|\psi (k_{1},k_{2})\right|^{2}$.

Explicit calculation based on \eqref{eq:ent-xi}, \eqref{eq:Pk1}, and \eqref{eq:Pk2} of the slices through the origin of $\bar{P}(k_{1},k_{2})$ gives the following conditional distributions (for economy of notation, we leave implicit the condition $s_{\pm}^{\perp}=0$):
\begin{widetext}
\begin{align}
\bar{P}(s_{\pm}) & = \frac{1}{a\pi}e^{-\frac{s_{\pm}^{2}}{2a}}\Bigg\{
 B^2\left[\cos\left(\frac{s_{\pm}h_{1}^{2}}{\sqrt{h_{1}^{2}+h_{2}^{2}}}\right)\cos\left(\frac{s_{\pm}h_{2}^{2}}{\sqrt{h_{1}^{2}+h_{2}^{2}}}\right)\cos\xi\mp\sin\left(\frac{s_{\pm}h_{1}^{2}}{\sqrt{h_{1}^{2}+h_{2}^{2}}}\right)\sin\left(\frac{s_{\pm}h_{2}^{2}}{\sqrt{h_{1}^{2}+h_{2}^{2}}}\right)\sin\xi\right]^{2}\nonumber \\
 & \quad-\frac{1}{8}B^4 \left[1+e^{-2ah_{2}^{2}}\cos\left(2\xi\right)+\left[\cos\left(2\xi\right)+e^{-2ah_{2}^{2}}\right]\cos\left(\frac{2s_{\pm}h_{1}^{2}}{\sqrt{h_{1}^{2}+h_{2}^{2}}}\right)\right] \nonumber \\
 & \qquad\qquad\times
\left[1+e^{-2ah_{1}^{2}}\cos\left(2\xi\right)+\left[\cos\left(2\xi\right)+e^{-2ah_{1}^{2}}\right]\cos\left(\frac{2s_{\pm}h_{2}^{2}}{\sqrt{h_{1}^{2}+h_{2}^{2}}}\right)\right]\nonumber\\
&+\frac{1}{8}B^4 \left[1+e^{-2ah_{2}^{2}}\cos\left(\frac{2s_{\pm}h_{1}^{2}}{\sqrt{h_{1}^{2}+h_{2}^{2}}}\right)\right]\left[1+e^{-2ah_{1}^{2}}\cos\left(\frac{2s_{\pm}h_{2}^{2}}{\sqrt{h_{1}^{2}+h_{2}^{2}}}\right)\right]
\Bigg\} ,
\end{align}
\end{widetext}
where we have applied an alternative method by \cite{Peled2020} to take the added and subtracted terms with the same coefficient $B^4(\xi)$ instead of using $B^4(\xi=\frac{\pi}{4})$ for the added term.
As a consequence of the addition and subtraction of different probability distributions, the resulting complicated quantum interference patterns can no longer be described with only two envelopes. Instead, detailed mathematical analysis shows that there is a complicated interplay between three distinct envelopes obtained with the following substitutions:
$\textrm{env}{}^{-}(s_{\pm})$ is obtained by setting $\frac{s_{\pm}h_{1}^{2}}{\sqrt{h_{1}^{2}+h_{2}^{2}}}\to\frac{\pi}{4}$ and $\frac{s_{\pm}h_{2}^{2}}{\sqrt{h_{1}^{2}+h_{2}^{2}}}\to\frac{\pi}{4}$;
$\textrm{env}{}^{+}(s_{\pm})$ is obtained by setting $\frac{s_{\pm}h_{1}^{2}}{\sqrt{h_{1}^{2}+h_{2}^{2}}}\to\frac{\pi}{4}$ and $\frac{s_{\pm}h_{2}^{2}}{\sqrt{h_{1}^{2}+h_{2}^{2}}}\to-\frac{\pi}{4}$;
and
$\textrm{env}{}^{0}(s_{\pm})$ is obtained by setting $\frac{s_{\pm}h_{1}^{2}}{\sqrt{h_{1}^{2}+h_{2}^{2}}}\to\frac{\pi}{2}$ and $\frac{s_{\pm}h_{2}^{2}}{\sqrt{h_{1}^{2}+h_{2}^{2}}}\to\frac{\pi}{2}$.

To compute the visibilities $\mathcal{V}\left(s_{\pm}\right)$, one needs to consider two cases:
if $h_1 \neq h_2$, the visibilities $\mathcal{V}\left(s_{\pm}\right)$ are computed from the pair of envelopes $\textrm{env}{}^{+}(s_{\pm})$ and $\textrm{env}{}^{-}(s_{\pm})$; whereas
if $h_1 = h_2$, the visibilities $\mathcal{V}\left(s_{\pm}\right)$ are computed from the pair of envelopes $\textrm{env}{}^{0}(s_{\pm})$ and $\textrm{env}{}^{-}(s_{\pm})$.
Then, the two-particle visibility of the ``corrected'' distribution becomes
\begin{equation}
F= \max\left[\mathcal{V}\left(s_{+}\right), \mathcal{V}\left(s_{-}\right) \right] .
\end{equation}
In the limit of infinite Gaussian precision $a\to\infty$, perfect complementarity is achieved only in the case when $h_1 \neq h_2$ :
\begin{equation}
\lim_{a\to\infty} \left(V^{2}+F^{2} \right) = \cos^{2}\left(2\xi\right)+\sin^{2}\left(2\xi\right) = 1.
\end{equation}
In the case when $h_1 = h_2$, one arrives only at an inequality as shown in Fig.~\ref{fig:7},
\begin{equation}
V^{2}+F^{2} \leq 1 .
\end{equation}

One drawback of determining the two-particle visibility from the ``corrected'' distribution is the appearance of three envelopes due to complicated interference patterns.
It should be noted that $\textrm{env}{}^{-}(s_{+})$ acts as a lower envelope when $\xi \in (0,\frac{\pi}{2})$ and as an upper envelope when $\xi \in (\frac{\pi}{2},\pi)$. Conversely, $\textrm{env}{}^{-}(s_{-})$ acts as an upper envelope when $\xi \in (0,\frac{\pi}{2})$ and as a lower envelope when $\xi \in (\frac{\pi}{2},\pi)$.
Analogously, $\textrm{env}{}^{+}(s_{+})$ acts as an upper envelope when $\xi \in (0,\frac{\pi}{2})$ and as a lower envelope when $\xi \in (\frac{\pi}{2},\pi)$. Conversely, $\textrm{env}{}^{+}(s_{-})$ acts as a lower envelope when $\xi \in (0,\frac{\pi}{2})$ and as an upper envelope when $\xi \in (\frac{\pi}{2},\pi)$. The envelope $\textrm{env}{}^{0}(s_{\pm})$ lies always between $\textrm{env}{}^{+}(s_{\pm})$ and $\textrm{env}{}^{-}(s_{\pm})$, except at the extreme values $\xi=0+n\frac{\pi}{2}$ when all three envelopes coincide with each other.
When $h_1 \neq h_2$, the slice distributions $\bar{P}(s_\pm)$ are bounded by the envelopes $\textrm{env}{}^{+}(s_{\pm})$ and $\textrm{env}{}^{-}(s_{\pm})$. Letting $h_2$ approach $h_1$ (or vice versa) creates an interference effect so that the central part of $\bar{P}(s_\pm)$ around $s_\pm = 0$ becomes bounded between $\textrm{env}{}^{0}(s_{\pm})$ and $\textrm{env}{}^{-}(s_{\pm})$, while leaving the outer tails of $\bar{P}(s_\pm)$ still located between $\textrm{env}{}^{+}(s_{\pm})$ and $\textrm{env}{}^{-}(s_{\pm})$. At the end of the transition $h_2 \to h_1$, when the exact equality $h_1 = h_2$ is reached, all of $\bar{P}(s_\pm)$ is bounded between $\textrm{env}{}^{0}(s_{\pm})$ and $\textrm{env}{}^{-}(s_{\pm})$. Because in real-world setups $h_1$ and $h_2$ can never be perfectly equal, measuring $F$ will always be confounded to some degree by the described transitioning from $\textrm{env}{}^{+}(s_{\pm})$ to $\textrm{env}{}^{0}(s_{\pm})$. In contrast, measuring $D$ is straightforward because the interference in the original ``uncorrected'' $P(s_\pm)$ is simple and involves only two envelopes.

Another drawback to measuring $F$ from the conditional distributions $P(s_\pm)$ is the tiny probability of postselecting $s_{\pm}^{\perp}=0$. This means that a large number of unsuccessful postselections need to be discarded from analysis.
In contrast, measuring $D$ from unconditional distributions $P(s_\pm)$ discards no experimental data and extracts the two-particle visibility with a smaller overall number of measured entangled pairs.

\section{Slice distributions and marginal distributions}
\label{app}

Throughout this work, we have analyzed the geometric properties of two-dimensional probability
distributions such as $P(k_{1},k_{2})$, which depend on two independent variables, $k_{1}$ and $k_{2}$.
The two main operations of interest for producing one-dimensional distributions from a given two-dimensional probability
distribution are \emph{slicing} or \emph{marginalization}.

A synopsis of the main differences between slice distributions and marginal distributions is as follows:

The \emph{slice distribution} is a one-dimensional conditional distribution in which the second variable is fixed to a specific value. Hence, the slice distribution is not normalized to~1. Because the use of integration is not required at all, consideration of the Jacobian is not needed after the change of basis. The use of the Dirac $\delta$ function for substitutions only complicates the math presentation.

The \emph{marginal distribution} is a one-dimensional unconditional distribution in which the second variable is not fixed and can be any value. Hence, the marginal distribution is normalized to~1. Because integration is required over the second variable, consideration of the Jacobian is needed after the change of basis. The use of the Dirac $\delta$ function simplifies the math presentation.

The meaning of the above summaries is unpacked in the following explicit definitions.

\begin{defn}
(Slice of two-dimensional probability distribution) The slice of two-dimensional probability distribution~$P(k_{1},k_{2})$ is
a one-dimensional probability distribution that is a function of only
one independent variable, e.g., $k_{1}$ when the second variable {is
fixed to a specific value}, e.g., $k_{2}=0$. Exactly because the second
variable is fixed to a specific value, the slice of a two-dimensional
distribution is referred to as a {conditional distribution}. In other
words, the two concepts {slice} and {conditional}
distribution are equivalent and can be used interchangeably. Furthermore,
integration with respect to the first variable, e.g., $k_{1}$ does
not give unit probability, but rather gives the probability density
for occurrence of the fixed outcome for the second variable, e.g.,
$\int_{-\infty}^{\infty}P(k_{1},k_{2}=0)dk_{1}=P(k_{2}=0)$.
\end{defn}

\begin{defn}
(Rotated slice) To cut a rotated slice parallel to an arbitrary $k_{u}$
axis through the two-dimensional distribution $P(k_{1},k_{2})$, one
needs to change basis from $k_{1},k_{2}$ to $k_{u},k_{v}$ and then
fix the value of the orthogonal $k_{v}$ variable (i.e., the second
variable). The change of basis is given by the transformation
\begin{equation}
\left(\begin{array}{c}
k_{u}\\
k_{v}
\end{array}\right)=\left(\begin{array}{cc}
\cos\varphi  & \sin\varphi \\
-\sin\varphi  & \cos\varphi
\end{array}\right)\left(\begin{array}{c}
k_{1}\\
k_{2}
\end{array}\right) .
\end{equation}
The inverse transformation is
\begin{equation}
\left(\begin{array}{c}
k_{1}\\
k_{2}
\end{array}\right)=\left(\begin{array}{cc}
\cos\varphi  & -\sin\varphi \\
\sin\varphi  & \cos\varphi
\end{array}\right)\left(\begin{array}{c}
k_{u}\\
k_{v}
\end{array}\right) .
\end{equation}
In other words, simple substitution in $P(k_{1},k_{2})$ of the following
identities,
\begin{eqnarray}
k_{1} & = & k_{u}\cos\varphi -k_{v}\sin\varphi , \nonumber\\
k_{2} & = & k_{u}\sin\varphi +k_{v}\cos\varphi , \label{eq:C3}
\end{eqnarray}
followed by fixing numerically the value of $k_{v}$, e.g., $k_{v}=0$,
will produce a {conditional} distribution of $k_{u}$ that
is a {rotated slice} of the two-dimensional distribution
$P(k_{1},k_{2})$. It should be noted that {no integration
is required} at all, only substitution based on mathematical equality.
\end{defn}

\begin{defn}
(Dirac delta function) One of the mathematical properties of the Dirac
$\delta$ function is that it allows {use of integration}
as a fancy way to {perform substitutions}. In particular,
if one has a function $f(k_{1},k_{2})$ in which one wants to fix
the $k_{1}$ value to a specific constant, e.g., $k_{1}=0$ thereby
obtaining $f(0,k_{2})$, it is possible to use a {single
integral} as follows:
\begin{equation}
\int_{-\infty}^{\infty}f(k_{1},k_{2})\delta(k_{1}-0)dk_{1}=f(0,k_{2}) .
\end{equation}
However, one can also use the Dirac $\delta$ function to simply {rename}
the variable $k_{1}$ into another letter, e.g., $k_{1}\to k_{s}$,
with exactly the same integral formula
\begin{equation}
\int_{-\infty}^{\infty}f(k_{1},k_{2})\delta(k_{1}-k_{s})dk_{1}=f(k_{s},k_{2}) .
\end{equation}
Therefore, it is in general incorrect to think of the integral of the Dirac
$\delta$ function as fixing the value of $k_{1}$, but rather as
{replacing} $k_{1}$ with something else, either {variable}
(``renaming'') or {constant} (``fixing'').
\end{defn}

\begin{defn}
(Marginal distribution) The {marginal} distribution $P(k_{1})$
obtained from the two-dimensional distribution $P(k_{1},k_{2})$ is
the {unconditional} distribution obtained by integration
over the second variable $k_{2}$ as follows:
\begin{equation}
P(k_{1}) = \int_{-\infty}^{\infty}P(k_{1},k_{2})dk_{2} .
\end{equation}
This {is not a slice} of $P(k_{1},k_{2})$ but a normalized
probability distribution for $k_{1}$ such that the second variable
$k_{2}$ is not fixed and can take any value. In other words, integration
of $P(k_{1})$ over $k_{1}$ returns the probability that $k_{2}$
will take any value at all, which is $1$ (because $k_{2}$ must have
some value).
\end{defn}

\begin{defn}
(Rotated marginal distribution) The rotated marginal distribution is obtained from the two-dimensional distribution $P(k_{1},k_{2})$ by integration along an arbitrary rotated axis $k_{v}$ \cite{Temme1987,Deans1983}.
The resulting distribution from the marginalization is a function of the orthogonal
variable $k_{u}$, which is renamed to $k_s$ using integration of the Dirac $\delta$ function,
\begin{equation}
P(k_{s})=\int_{-\infty}^{\infty}\int_{-\infty}^{\infty}P(k_{1},k_{2})\delta\left(k_{s}-k_{1}\cos\varphi -k_{2}\sin\varphi \right)dk_{1}dk_{2} . \label{eq:rmd-1}
\end{equation}
It is worth emphasizing that we treat $\varphi$ as being fixed to a specific value. Furthermore, we use rotated Cartesian coordinates instead of polar coordinates.
The {change of variables} from $k_{1},k_{2}$ to rotated
$k_{u},k_{v}$ in the {double integral} using transformation \eqref{eq:C3}
requires consideration of the {Jacobian}
\begin{equation}
J=\left|\begin{array}{cc}
\cos\varphi  & -\sin\varphi \\
\sin\varphi  & \cos\varphi
\end{array}\right|=\cos^{2}\varphi +\sin^{2}\varphi =1,
\end{equation}
which relates the differentials
\begin{equation}
dk_{1}dk_{2}=Jdk_{u}dk_{v}.
\end{equation}
Changing the variables in explicit algebraic steps gives
\begin{widetext}
\begin{align}
P(k_{s}) & =\int_{-\infty}^{\infty}\int_{-\infty}^{\infty}P(k_{1},k_{2})\delta\left(k_{s}-k_{1}\cos\varphi -k_{2}\sin\varphi \right)dk_{1}dk_{2}\\
 & =\int_{-\infty}^{\infty}\int_{-\infty}^{\infty}P(k_{u},k_{v})\delta\left(k_{s}-k_{u}\right)Jdk_{u}dk_{v}\label{eq:16}\\
 & =\int_{-\infty}^{\infty}P(k_{s},k_{v})dk_{v} . \label{eq:rmd-2}
\end{align}
\end{widetext}
From the last integral \eqref{eq:rmd-2} it can be seen that the rotated marginal distribution is not a slice distribution
because $k_{v}$ is not fixed to a specific value, but rather $k_{v}$
is integrated over. Performing the first integral \eqref{eq:16} used
the Dirac $\delta$ function to {rename} one of the variables
$k_{u}$ into $k_{s}$. This first integration does not produce a
slice because $k_{s}$ is not a constant. The second integration over
$k_{v}$ is the essential one that performs the marginalization. Note that if $k_{s}$ is assumed to be a constant,
e.g., $k_{s}=0$, then the result from the marginalization will not
be a distribution, but the {value at a single point}, e.g., $P(k_{s}=0)$
of the marginal distribution. One can say that formula \eqref{eq:rmd-2}
is a somewhat simpler way to define the rotated marginal distribution, namely, one has to specify a rotated axis~$k_{u}=k_{s}$
and then integrate over the orthogonal axis $k_{v}$. This, however,
requires additional specification in the text of the rotation matrix by providing
the angle $\varphi $ separately from the integral formula. The fancy
definition (\ref{eq:rmd-1}) involving the Dirac $\delta$ function
has the advantage that it already contains the rotation angle $\varphi$ displayed inside
the math expression \cite{Temme1987,Deans1983}.
\end{defn}

\section{Fitting of envelopes from empirical data}
\label{app-2}

The probability distribution $P(k_{1})$ given by \eqref{eq:Pk1} has the following upper and lower envelopes:

\begin{align}
\textrm{env}^{+}(k_{1}) & = e^{-\frac{k_{1}^{2}}{2a}} \frac{B^{2}}{2\sqrt{2a\pi}}\left(1+ e^{-2ah_{2}^{2}}\right)\left[1+\cos\left(2\xi\right)\right]
\nonumber\\
& = e^{-\frac{k_{1}^{2}}{2a}} A_{+} ,
\label{eq:env-k1+}
\end{align}
\begin{align}
\textrm{env}^{-}(k_{1}) & = e^{-\frac{k_{1}^{2}}{2a}} \frac{B^{2}}{2\sqrt{2a\pi}}\left(1- e^{-2ah_{2}^{2}}\right)\left[1-\cos\left(2\xi\right)\right]
\nonumber\\
& = e^{-\frac{k_{1}^{2}}{2a}} A_{-} .
\label{eq:env-k1-}
\end{align}
Because we are interested in the limit $a\to\infty$ we assume that $a$ is known in advance and fixed at the maximal value that is feasible under the current quantum technology. Since the two amplitudes $A_{+}$ and $A_{-}$ are constants independent of $k_1$, and we know that both envelopes are Gaussians of the form $e^{-\frac{k_{1}^{2}}{2a}}$, it is straightforward to find the best (least-squares) linear fit for $A_{+}$ using only the data points corresponding to local maxima, or $A_{-}$ using only the data points corresponding to local minima. The visibility will then be
\begin{equation}
\mathcal{V}(k_1)= \frac{A_{+} - A_{-}}{A_{+} + A_{-}} .
\end{equation}
Fitting based on empirical data for the other visibilities is analogous.

\bibliographystyle{apsrev4-1}
\bibliography{references}

\end{document}